\documentstyle[aps,prl,multicol,psfig]{revtex}
\tighten
\begin{document}

%
%
\title{Non-Fermi-liquid behavior and
$d$-wave superconductivity near the charge-density-wave
quantum critical point}
\author{C. Castellani, C. Di Castro, and M. Grilli}
\address{Istituto di Fisica della Materia e
Dipartimento di Fisica, Universit\`a di Roma ``La Sapienza'',\\
Piazzale A. Moro 2, 00185 Roma, Italy}
\maketitle
%
%
\begin{abstract} 
A scenario is presented, in which the presence of a quantum
critical point due to formation of incommensurate charge density
waves accounts for the basic features of the high temperature
superconducting cuprates, both in the normal and in the superconducting
states. Specifically, the singular interaction arising close to
this charge-driven quantum critical point gives rise to the non-Fermi liquid 
behavior universally found at optimal doping. This interaction is
also responsible for $d$-wave Cooper pair formation with a superconducting
critical temperature strongly dependent on doping in the overdoped
region and with a plateau in the optimally doped region.
In the underdoped region a temperature dependent pairing
potential favors local pair formation
without superconducting coherence, with a peculiar temperature 
dependence of the pseudogap
and a non-trivial relation between the pairing temperature
and the gap itself. This last 
property is in good qualitative agreement with 
so far unexplained features of the experiments.
\end{abstract}
%
%

%
%

\begin{multicols}{2}

\section{INTRODUCTION}
Together with their large superconducting critical temperatures,
the cuprates display several anomalous normal state properties
which cannot be described in terms of
the standard Fermi liquid (FL) theory. 
One possible explanation for these anomalous properties 
of the normal phase is that the
low dimensionality of such highly anisotropic systems and their
correlated nature are at the origin of a breakdown of the FL. 
FL theory indeed breaks down in a variety of physical situations as, for 
instance, in quasi-one dimensional conductors. 
The one-dimensional metallic phase is described 
by the so-called Luttinger Liquid theory with
no quasi-particle weight at the Fermi surface \cite{sol}. 
The assessed breakdown of FL theory in $d=1$ suggested
the very intriguing theoretical question of a non-Fermi Liquid  
metallic behavior in two-dimensional
electron systems as an extension of the one-dimensional 
case \cite{and}. However, it was recently found
\cite{clc} that,
for non-singular  interactions (involving small momentum transfer) 
Luttinger behaviour is strictly confined to one-dimensional metals. 
Only sufficiently singular
(stronger than the purely coulombic interaction)
long-range forces give non-FL  at low  energy above 
one dimension \cite{mac,bw}. The anomalous properties
would then arise as a consequence of singular scattering processes at low
energy.

Singular scattering can be due to gauge 
field fluctuations \cite{nagaosa}, which
arise by implementing the resonating-valence-bond idea in the
t-J model. An alternative point of view is related to the existence of a 
quantum critical point (QCP), where critical fluctuations 
can mediate singular interactions between the quasiparticles,
providing at the same time a strong pairing mechanism.
The proximity to the critical point at zero temperature
is naturally characterized by the absence of any energy scale
besides the temperature itself. This would agree with 
what can be inferred from the many
anomalous features of the normal state at optimal doping.
Various proposals have been
put forward on the possible nature of the QCP, ranging from 
a magnetic mechanism \cite{MBP,pines,pinesQCP}
to charge-transfer \cite{varma}
or to incommensurate charge-density-wave (ICDW) \cite{CDG,PCDG}.

In the antiferromagnetic (AF) QCP proposal 
a difficulty arises. Specifically, the most evident
features of non-FL behavior in the normal phase and the largest
superconducting critical temperature occur at optimal doping
which would be in the quantum disordered region
far away from the position of the AF transition. To explain this
behavior a mechanism (substantial vertex corrections) has to 
be advocated \cite{chubukov} to suppress the effect of critical 
fluctuations, which would otherwise be the strongest at low doping
just above the AF transition. Besides the fact that in the 
disordered regime the depression of the effective interaction
due to vertex correction is not established,
it would be hard to understand why at optimal doping
the best non-FL behavior
always occurs for all the classes of materials. 
Indeed within this scenario optimal doping is a quite generic
point of the quantum disordered region with a finite
energy scale.

The peculiarities of optimal doping make this point a natural
candidate where to place a QCP relevant for the
superconducting and the non-FL properties of the cuprates.

As far as the nature of this
critical point is concerned, it seems quite likely
that charge degrees of freedom should play the major role,
since the disordered region of this QCP coincides with the
highly metallic overdoped regime. In this context, besides the appealing
but rather exotic proposal of a simmetry breaking related to
persistent charge-transfer currents \cite{varma}, we believe that an
ICDW-QCP has strong support from both the theoretical and 
the experimental point of view.
The existence of an ICDW-QCP
is not alternative to the existence of 
an AF-QCP and the two QCP's control the behavior of the system
at different doping.  
The ICDW-QCP sets up the maximum superconducting critical temperature 
and can constitute the substrate to substain AF fluctuations 
far away from the AF ordered phase, by allowing for 
hole-rich and hole-poor ``stripes''.

After phase separation (PS) was shown to be present in the phase
diagram of the t-J model \cite{emery,marder}, we pointed out
that PS commonly occurs in models with short
range interactions\cite{GRCDK1}-\cite{BTGD}, 
provided a strong local
$e$-$e$ repulsion inhibits the stabilizing
role of the kinetic energy. We therefore stressed
that PS and superconductivity are
related phenomena irrespective of the nature
of the short-range interaction (magnetic, excitonic, 
phononic,...) \cite{notanumer}.

Emery and Kivelson \cite{emerykivelson} suggested that,
although long-range Coulomb (LRC)
forces spoil PS as a static thermodynamic
phenomenon, the frustrated tendency towards PS
may still be important and give rise
to large-amplitude charge collective density fluctuations.
Approaching the problem within a coarse-grained model,
they suggested that these
fluctuations may be responsible for the anomalous
behaviour of the normal phase.
A static pseudospin formulation \cite{low} of these ideas
showed the formation of a phase with hole-rich and hole-poor
stripes. This latter result is on the same line of the finding of Refs.
\cite{RCGBK,CDG,BTGD}, where an ICDW phase was shown
to arise in models where PS is spoiled by LRC forces.
Our finding is then that in all these models there
exists a QCP associated to the formation of ICDW. 
Near this QCP the dynamic effective interaction
between quasiparticles has a  singular behavior \cite{CDG}, 
strongly affecting the single-particle 
properties and the transport scattering time.
In the Cooper channel the same singular scattering
provides a strong pairing mechanism with an anisotropic
order parameter of $d$-wave symmetry \cite{BTGD,PCDG}.

Several experimental findings provide support for the existence of a
QCP at (or near) optimal doping. This is found in 
recent transport experiments \cite{boebinger} 
in ${\mathrm La_{2-x}Sr_xCuO_4}$ (LSCO), with high
magnetic fields, which allow to access the normal phase hidden
by superconductivity, and assess the presence of a metal-insulator
transition ending at optimal doping at $T=0$. Indications in the
same sense are provided by neutron scattering \cite{aeppli}
revealing a huge increase of a magnetic dynamical correlation
length in nearly optimally doped LSCO. Qualitative changes of
behavior at optimal doping are also detected by 
optical spectroscopy \cite{puchkov}, NMR \cite{jullien}, 
susceptibility \cite{batlogg},
neutron scattering \cite{rossat}, 
photoemission \cite{marshall,harris,campuzano},
specific heat \cite{loram}, thermoelectric power \cite{zhou},
Hall coefficient \cite{hwang}, 
resistivity \cite{batlogg,ito,boebinger}.
It is also suggestive that several quantities
(resistivity, Hall number, uniform susceptibility) display
a scaling behavior with a typical energy scale, which 
vanishes at optimal doping \cite{johnston,nakano,wuyts}.

Many indications exist that the above QCP involves charge ordering.
In Ref. \onlinecite{boebinger} the metal-insulator transition at 
T=0 occurs with a high value of $k_Fl$ 
(clean limit) suggesting that some charge
ordering underlies the insulating behavior of the underdoped
LSCO samples. A direct observation of charge-driven ordering
was possible by neutron scattering
\cite{tranquada1,tranquada2,tranquada3},
in ${\mathrm La_{1.48}Nd_{0.4}Sr_{0.12}CuO_4}$ 
where the related Bragg peaks were detected. For this specific
compound the low temperature tetragonal lattice structure 
pins the CDW and gives static order and 
semiconducting behavior (see also the case of 
${\mathrm La_{1.88}Ba_{0.12}CuO_4}$).
Increasing the Sr content at fixed Nd concentration, the pinning
effect is reduced leading to metallic and superconducting behavior.
In this latter case, the existence of dynamical
ICDW fluctuations is suggested by the presence of dynamical incommensurate
spin scattering, although 
the charge peaks are too weak to be observed. 
In this regard, also the 
${\mathrm La_{2-x}Sr_xCuO_4}$ is expected to display dynamical charge
fluctuations with doping-dependent spatial modulation as
indeed observed in the magnetic scattering \cite{yamada}.
ICDW have been proposed from extended X-ray absorption
fine structure (EXAFS) experiments both in optimally
doped LSCO \cite{bianconi1} and ${\mathrm Bi_2Sr_2CaCu_2O_{8+x}}$
(Bi-2212) \cite{bianconi2}.
Superstructures have also been detected in Bi-2212 from 
X-ray diffraction \cite{bianconi3}. 

In the next section we will consider the singular interactions
arising in the proximity of charge instabilities. In particular
we will consider an ICDW instability, which is present in strongly
correlated models due to the interplay between PS and LRC forces.
This instability is found \cite{CDG,BTGD} in specific models
at T=0 starting from a uniform FL phase
describing the low-temperature overdoped phase of
the superconducting cuprates. Once the ICDW instability
will be shown to occur, the challenging task remains of 
providing a complete description of the systems at finite
temperature and in the underdoped phase,
where superconductivity, (dynamical) ICDW order and magnetism
interplay. While a full theoretical understanding of this
latter phase is still missing, in Section III
we will provide a general scenario for the T vs. doping $\delta$
phase diagram of the cuprates based both on theoretical
results and on experimental evidences. 
Our conclusions are presented in Section IV.

\section{Singular scattering close to charge instabilities}

The evaluation of the density-density correlation function
$$
\chi (q,\omega) \equiv \langle n(q,\omega) 
n(-q,-\omega)\rangle
$$
provides information on the (charge) stability. In particular
a divergence in the static correlation function
$\chi (q,\omega=0)$ signals the  occurrence of
PS (at $q \to 0$) 
or of CDW instabilities (at finite $q$'s).
A complete investigation of the static and dynamical properties of the
infinite-U Hubbard-Holstein 
model together with the analysis of its stability 
was carried out in a previous work \cite{CDG,BTGD}.
Here we just mention that, within a large-N slave-boson formalism,
this model displays a phonon-driven 
charge instability even for rather small values of the
electron-phonon coupling $g$ \cite{GC}.
In the absence of LRC forces  the PS instability occurs
before any other finite $q$ instability. The introduction 
of LRC forces eliminates the $q=0$ divergence in the 
static correlation function
always giving rise to ICDW. The finite critical $q_c$,
in this case, is not related to any pseudonesting
of the Fermi surface. $q_c$ is determined by
the momentum dependence of the
(divergent) static correlation function with
only short-range forces and by the strength $V_C$, which parametrize
the LRC forces according to 
$$
V_{LR}(q)={V_C \over (a_\perp/a_\parallel) \sqrt{\epsilon_\parallel /
\epsilon_\perp}}{1\over q}.
$$
The momenta are in units of the inverse planar lattice spacing $a_\parallel$,
$a_\perp$ is the interplane distance and 
$\epsilon_{\parallel} $ and $\epsilon_{\perp}$ are 
the corresponding dielectric constants. 

A divergent scattering amplitude 
between quasiparticles $\Gamma (q,\omega)$ will follow from
a divergent correlation function $\chi$.
Indeed the interaction between the quasiparticles
is mediated by  the exchange of bosonic degrees of freedom,
e.g. phonons and slave-boson fields accounting for the strong local
repulsion U. These bosonic excitations have 
a singular propagator, which
enters in the expressions of both $\Gamma$ and $\chi$
establishing a clear connection between the charge instability 
and the singular quasiparticle scattering. 

Near the PS instability ($V_C=0$), the anomalous behavior 
of $\Gamma$ is identified to be of the form \cite{CDG}
\begin{equation}
\Gamma (q,\omega) \approx 
\tilde{U} - {V \over q^2 - i\omega {\gamma \over q} +\kappa^2} .
\label{fitgamsr}
\end{equation}
$\tilde{U}$ describes the (almost momentum-independent, i.e.
local) residual repulsion mediated by the slave bosons: Within
the large-N slave-boson formalism, the infinite U repulsion between the 
bare fermions, is reduced to a rather weak
residual repulsion between the Fermi-liquid quasiparticles.
$\gamma$ is a damping coefficient.
The mass term $\kappa^2=a(\delta-\delta_c)$ vanishes linearly
when, for a given $g$, the instability takes place
at the critical doping $\delta_c=\delta_c(g)$.
$\kappa$ can be interpreted as the inverse
correlation length for the density fluctuations,
$\kappa = \xi^{-1}$.
It is worth noticing that the singular part of the effective interaction
in Eq. (\ref{fitgamsr}) has the same functional form as
the scattering amplitude mediated by gauge fields \cite{nagaosa},
although its physical origine is obviously different.

The singular behavior
of $\Gamma_q=\Gamma(q \to 0,\omega=0)$ at the PS instability is
by no means surprising within a FL framework. Indeed, the
FL expression for the compressibility is $\chi_n=2\nu^*/\left(1+2\nu^*
\Gamma_\omega \right)$, where $\Gamma_\omega=
\Gamma_q/(1-2\nu^*\Gamma_q)$
is the standard dynamic ($\omega \to 0, q=0$) 
limit of the scattering amplitude $\Gamma (q, \omega)$
and $\nu^*$ is the quasiparticle density of states 
at the Fermi level.
This indicates that a divergent $\chi_n$,
when the quasiparticle mass remains 
finite ($\nu^*<\infty$), only happens when 
$2\nu^*\Gamma_\omega \to -1$ (Pomeranchuk criterion).
At the same time $\Gamma_q \to -\infty$.
 We point out here that the above arguments keep their
validity irrespective of the mechanism leading to PS. 
However PS is related to a first order transition and
the need of a Maxwell construction introduces in the phase diagram 
a cohexistence region embedding
the spinodal instability line. Except for the end critical point,
the compelled distance of the stable region from the instability
line may render the above mechanism for obtaining 
singular scattering non-generic.

We now proceed to the more likely situation which originates
from the presence of LRC forces. 
In this case the singular part of $\Gamma$
can be written as
\begin{equation}
\Gamma ({\mbox{\boldmath $q$}},\omega) \approx 
\tilde{U} - {1 \over 4} \sum_\alpha \frac{V}{\kappa^2+
\omega_{\mbox{\boldmath $q$}}^{\alpha}
- i\gamma \omega} 
\label{fitgamlr}
\end{equation}
where the sum is over the  four equivalent vectors of the CDW 
instability ${\mbox{\boldmath $q$}}^{\alpha}
=(\pm q_c,0),(0,\pm q_c)$ and 
$\omega_{\mbox{\boldmath $q$}}^{\alpha} =
2(2-\cos(q_x-q_x^{\alpha})-\cos(q_y-q_y^{\alpha}))$.
This expression is used to reproduce the behavior $\sim -1/(\kappa^2+
(q_x-q_x^{\alpha})^2+(q_y-q_y^{\alpha})^2)$ for
$q\rightarrow q^{\alpha}$ while mantaining the lattice periodicity. 

Also in this case a linear
behavior of the mass term $\kappa^2=a(\delta-\delta_c)$ was found.
For reasonable parameters 
(e.g., for LSCO systems we considered a first and next nearest
neighbor hopping
$t=0.5$ eV and $t'/t=-1/6$ respectively, 
$V_C=0.55$ eV, a dispersionless phonon with frequency
$\omega_0=0.04$ eV and electron-phonon coupling
$g=0.17$ eV) the instability first occurs at $\delta_c\approx 0.2$
with ${\mbox{\boldmath $q$}}_c\approx(\pm 0.28,\pm 0.86)$, or
${\mbox{\boldmath $q$}}_c\approx(\pm 0.86,\pm 0.28)$.
From our analysis of the infinite-U Hubbard-Holstein
model and of other models we found
that the rather large density of states near the $(\pm \pi,0)$
and $(0,\pm \pi)$ points tends to favor instabilities at or close
to the (1,0) or (0,1) directions.
However, as shown in Fig. 1, the scattering is 
quite strong, although non-singular, in all directions
for $\vert {\mbox{\boldmath $q$}} \vert \approx \vert 
{\mbox{\boldmath $q$}}_c \vert$.
The (almost) isotropic 
contribution to the static scattering amplitude is much less
fragile under doping variations, than the singular term itself.
The imaginary term in the denominators in the
r.h.s. of Eqs.(\ref{fitgamsr}) and (\ref{fitgamlr})
reproduces on a wide range of
transferred momenta $q$ the behaviour of the imaginary
part of the mean field fermionic
polarization bubble $Im\left[ \chi^0({\mbox{\boldmath $q$}},
\omega)\right] \propto
\omega /q$ at small $\omega $.  This indicates, that,
despite the complicated formal structure of the
scattering amplitude within the slave-boson
formalism, near the instability a simple RPA-like structure results 
in the final expression. This supports the idea that
the forms (\ref{fitgamsr})
and (\ref{fitgamlr}) are 
generic of PS or ICDW and not related to the specific mechanism,
which gives rise to the tendency towards phase separation.
\begin{figure}
\vspace{-1 truecm}
\centerline{\hbox{\psfig{figure=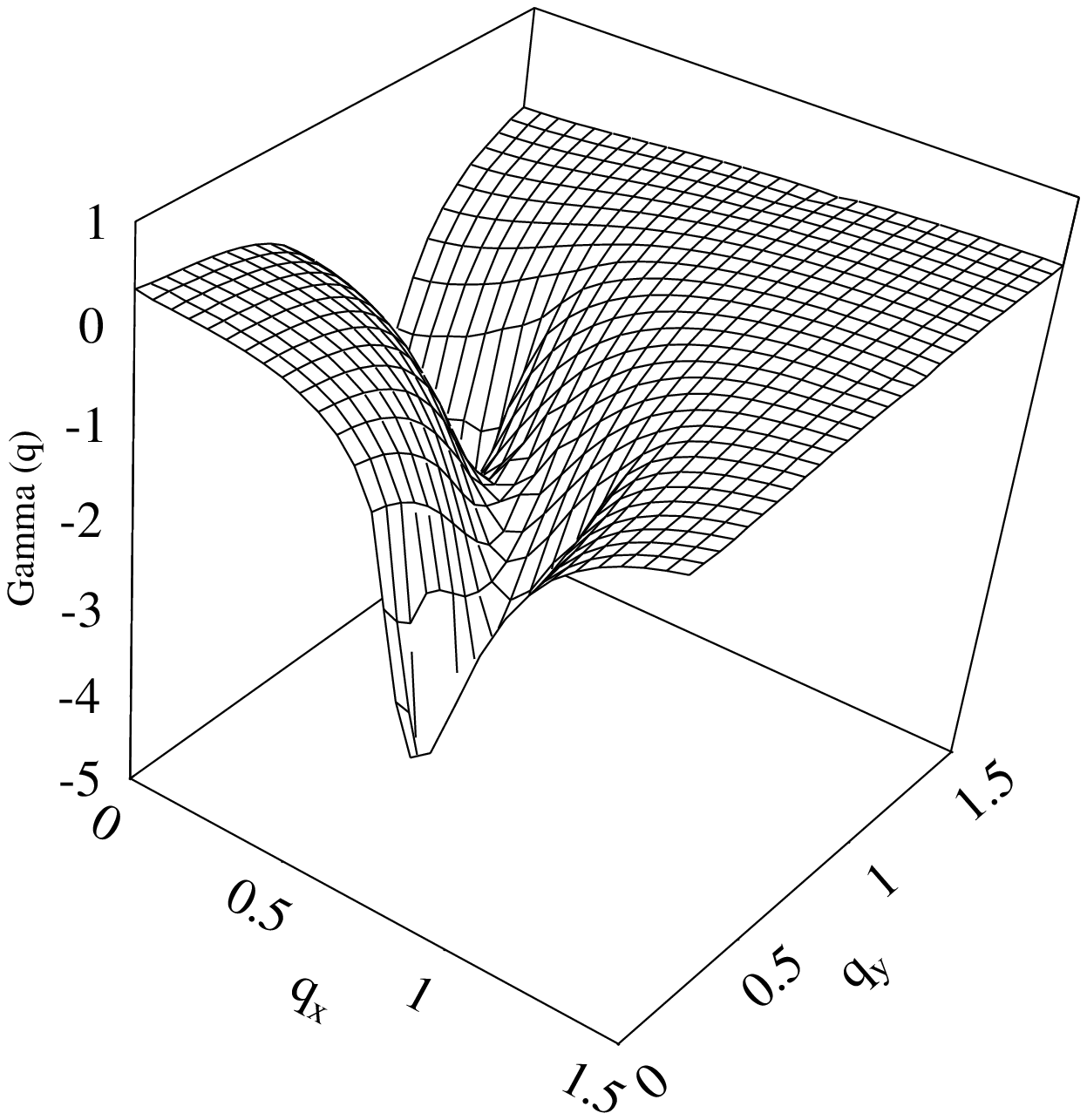,width=7.cm,angle=-90}}}
\vspace{-1 truecm}
{\small FIG. 1: Momentum dependence of the static 
scattering amplitude in the infinite-U Hubbard-Holstein model
for $t=0.5$ eV $t'=-1/6t$, $V_C=0.55$ eV, 
$\omega_0=0.04$ eV and $g=0.17$ eV at $\delta=0.195$.}
\label{FIG1}
\end{figure}

\section{The ICDW quantum critical point and the phase diagram}

In the previous section an ICDW 
instability was shown to occur at T=0 by decreasing doping
from an overdoped strongly correlated FL system. According
to the scenario outlined in Section I, this ICDW-QCP is the
crucial ingredient characterizing the physics of the cuprates
all over their metallic (under-, optimally, and over-doped)
regime. As seen in Sec. II, the ICDW-QCP is characterized by 
singular interactions, which in the RPA-like treatment
of the model assumed the form of Eq.(\ref{fitgamlr}).
This functional dependence is not strictly related to
the specific origine of the QCP, as also witnessed by its similarity
to the interactions proposed for systems near the 
AF-QCP \cite{MMP}, where the instability also occurs for
a finite value of $Q_{AF}=(\pi,\pi)$.
However Eq.(\ref{fitgamlr}) 
could depend from our approximate
(nearly-mean-field) treatment. Nevertheless the singular
nature of interactions mediated by critical fluctuations
is a sound consequence. For the sake of definiteness
and simplicity, we will assume the form (\ref{fitgamlr}) to be
generically valid and we will also assume a simple Gaussian behavior 
of the QCP.

\subsection{The ICDW-QCP in the absence of pairing}
From the theory of QCP's \cite{QCPH,QCPAF,QCP}, 
one can schematically draw Fig. 2, which would be valid {\it in the
absence of any superconducting pair formation}. 
\begin{figure}
\centerline{\hbox{\psfig{figure=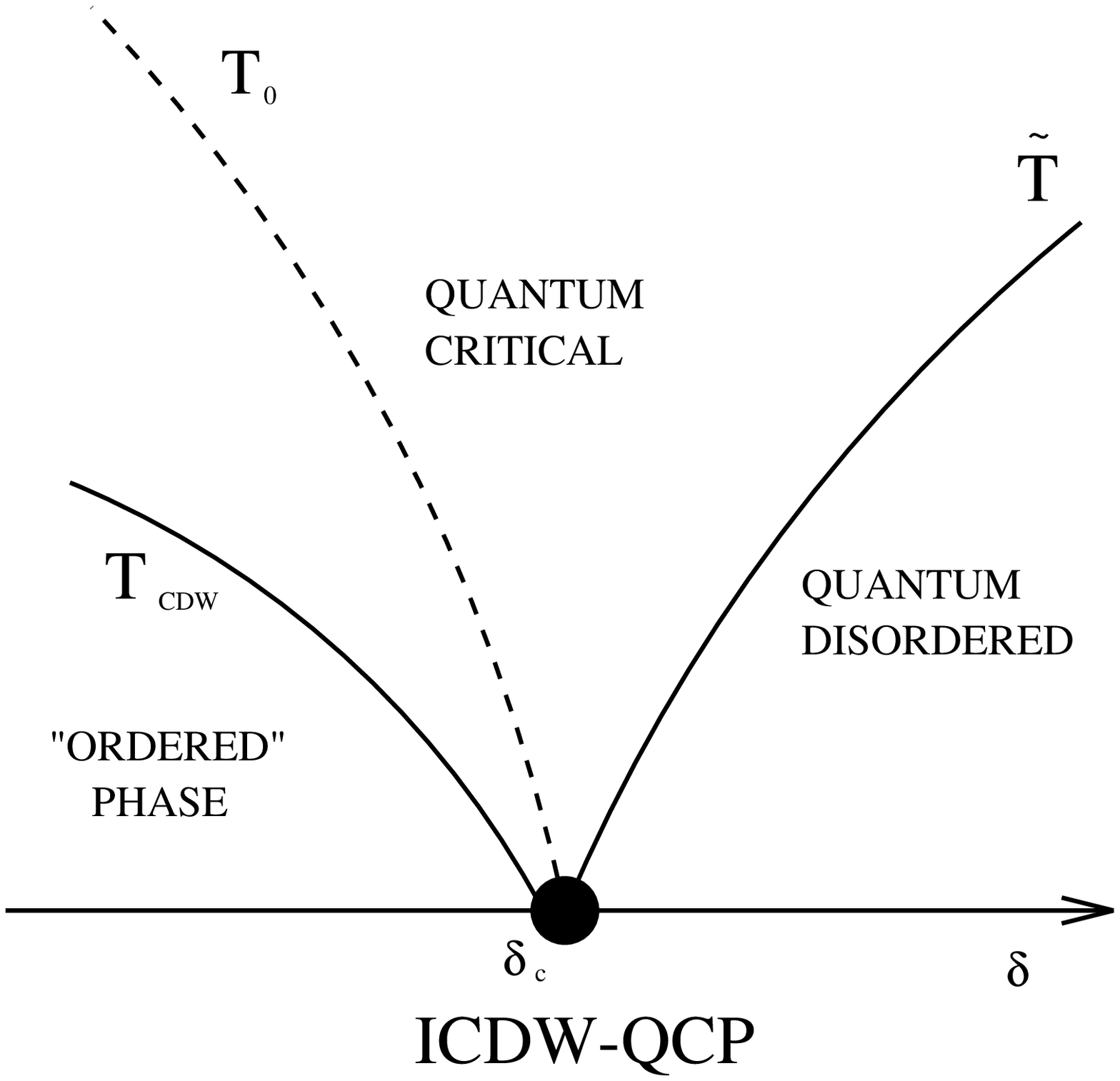,width=7.cm,angle=-90}}}
{\small FIG. 2: Schematic structure of the phase diagram
around the ICDW-QCP
in the absence of superconducting
pairing. On the right: Quantum disordered region
[$\xi^{-2}\approx (\delta-\delta_c)$]; In the middle: 
Quantum critical  (classical gaussian) region 
[$\xi^{-2}\approx T$]; On the left: Ordered ICDW phase.
The dashed line indicates the mean-field critical temperature line.
}
\label{FIG2}
\end{figure}
The overdoped region on the right corresponds to the
quantum disordered regime, where 
$\kappa^2 =\xi^{-2} \sim a(\delta-\delta_c)^{2\nu}$. 
Increasing the temperature we enter in the so-called
classical gaussian regime where
$\kappa^2$ becomes a function of $T$, 
$\kappa^2 \sim b T^{(d+z-2)/z}$. The crossover
occurs along the line 
$\tilde{T}=(a/b)^{z/(d+z-2)}
*(\delta-\delta_c)^{2\nu z/(d+z-2)}$, where $d$ is the spatial dimension, 
$z$ is the dynamical critical index. Roughly we can write 
\begin{equation}
\label{kmax}
\kappa^2= Max \left[ a(\delta-\delta_c)^{2\nu},b T^{(d+z-2)/z}\right]
\end{equation}
with $a$ and $b$
model-dependent positive constants, in order to represent the (much more 
complex) crossover of the actual $\kappa^2(\delta-\delta_c,T)$. 
The proper $z$ is $z=2$ for CDW as one sees
from  the fluctuation propagator.
In $d=2$ its value is however immaterial, since Eq.(\ref{kmax})
reduces to $\kappa^2=Max[a(\delta-\delta_c)^{2\nu},bT]$. 
As far as the index $\nu$ 
is concerned, since we are dealing with a QCP within the
classical gaussian approximation, we take $\nu=1/2$.

The region on the left would generically correspond to
the ordered ICDW phase occurring below a critical temperature
$T_{CDW}(\delta)$ starting from the QCP 
at $\delta_c(T=0)$.
The true critical line is depressed with respect to
its mean-field expression (sketched by the dashed line 
$T_0$ of Fig. 2).
When evaluated within a specific model,
$T_0$ is not only determined by the ${\cal O}(T^2)$
mean-field critical temperature for ICDW formation
in a metallic FL phase. It should also include the much
more important one-loop gaussian 
corrections (see Ref. \onlinecite{QCP}) accounting for the 
quantum dynamical reduction controlled by the proximity to the QCP.
The region between the two curves 
$T_0$ and $T_{CDW}$ is dominated by 
strong precursor effects. 
On the other hand, in $d=2$, the order parameter
strictly appears at T=0 only (in the clean limit).

The quantum disordered region on the right corresponds
to the overdoped region of the cuprates with FL behavior.
On other hand the classical gaussian region around
optimal doping is characterized by the absence
of energy scales, but the temperature. 
Here the best non-FL behavior is obtained.
In particular, with the scattering of the form 
given in Eq. (\ref{fitgamlr}), a linear-in-T resistivity
is expected in $d=2$, while for $d=3$, $\rho(T)\sim T^{3/2}$.
In Ref. \onlinecite{hlubina} the objection was raised
for magnetically mediated scattering, 
that only few "hot" points would feel strong scattering
contributing to the above behavior. Generically, 
all other points would contribute to the lower 
$T^2$ behavior. However, for ICDW, the fact that 
typical $q_c$ are fairly small and the strong isotropic
character of $\Gamma(q)$ shown in Fig. 1 
make this objection less relevant.

The scenario presented so far should find a 
physical correspondence whenever the superconducting pair
formation is forbidden. Indeed the transport experiments
in LSCO under strong magnetic field \cite{boebinger}
display a metal-insulator crossover, 
ending at T=0 at a QCP at optimal doping
(see Fig. 3 in Ref. \onlinecite{boebinger}).  
The experimental line separating the planar
metal from the insulator would correspond
in our picture to the ``true'' $T_{CDW}$
critical temperature as a function of doping.
We find quite remarkable that for filling close to the ``magic''
value 1/8, the experimental temperature for the insulator is
substantially higher (see Fig.1(a) of Ref. \onlinecite{boebinger})
than for nearby filling values.
This is consistent with the idea that commensuration effects
at this particular doping pin the thermal ICDW fluctuations
leading to a high $T_{CDW}$, strengthening the indications of a
charge-ordering phenomenon.

As far as $T_0$ is concerned, this temperature marks the
onset of the ICDW precursors and is characterized by 
a loss of spectral weight at low energies
giving rise to a uniform decrease of the density of states
near the Fermi energy. This would show up as the well known decrease
of the uniform magnetic susceptibility below a characteristic
temperature, which vanishes by approaching from below the
optimal doping \cite{johnston,nakano,wuyts}.
In underdoped ${\mathrm {YBa_2Cu_3O_{6+x}}}$ (YBCO)
compounds, this last temperature has also been 
put in correspondence \cite{ito,batlogg}
 with the temperature below which
the planar resistivity $\rho_{ab}$ deviates from its linear behavior,
while the interplane resistivity $\rho_c$
acquires a non metallic 
behavior \cite{takenaka,notarhoc}. This finding is then compatible 
with the further identification of our $T_0$ line
with the second line present in the phase diagram of LSCO in 
 high magnetic field. This second line
in Fig. 3 of Ref. \onlinecite{boebinger} separates a region at larger
doping where both $\rho_{ab}$ and $\rho_c$ have metallic
behavior from a region where $\rho_c$ increases with decreasing
temperature. Consistently with our scenario this crossover
line also ends at the QCP at optimal doping.

\subsection{The ICDW-QCP in the presence of pairing}
The above scenario is drastically modified, once superconducting
pairing is considered \cite{PCDG}. In particular it is found that the
singular interaction of Eq.(\ref{fitgamlr}) is also present
in the particle-particle channel, thus providing a strong
pairing mechanism
in the proximity of the critical point. 
Fig. 3 is a schematic representation of the phase diagram
near the ICDW-QCP by allowing for  superconducting pairing.
\begin{figure}
\centerline{\hbox{\psfig{figure=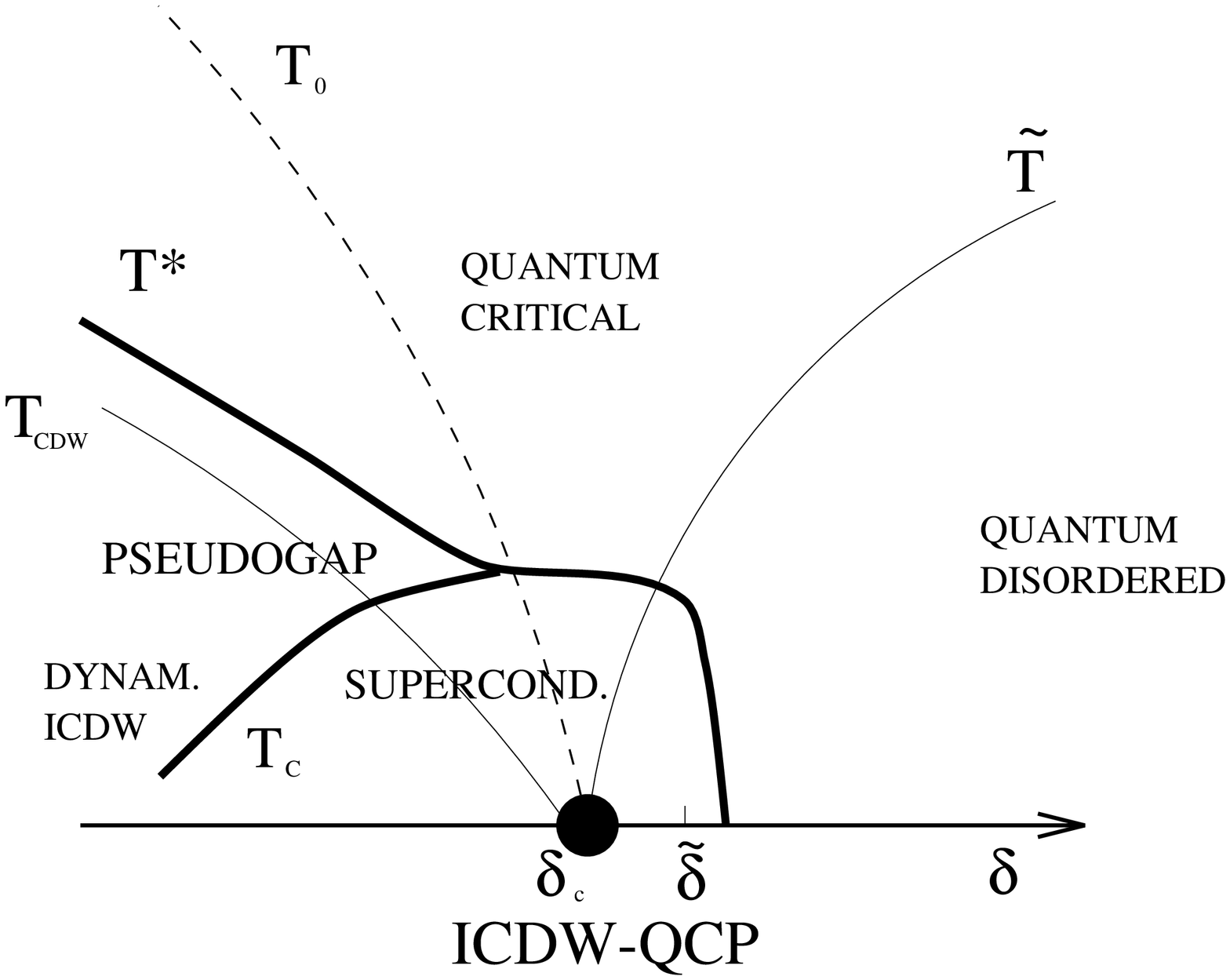,width=7.cm,angle=-90}}}
{\small FIG. 3: Schematic structure of the phase diagram
around the ICDW-QCP in the presence
of superconducting pairing. On the right: Quantum disordered region
[$\xi^{-2}\approx (\delta-\delta_c)$]; In the middle: 
Quantum critical (classical gaussian) region 
[$\xi^{-2}\approx T$]; On the left:  Dynamical ICDW phase.
The heavy line indicates the region of local (pseudogap)
or coherent (superconducting) pairing}
\label{FIG3}
\end{figure}
The most apparent and generic feature is that pairing
has $d$-wave symmetry and, being mediated
by an interaction  
rapidly varying with $\kappa^2$ [cf. Eq.(\ref{fitgamlr})],
strongly depends on temperature or doping.

We start discussing the overdoped quantum disordered regime.
Even in this uniform FL phase
the evaluation of a 
quantitatively reliable superconducting critical temperature is a
difficult task since the pairing is mediated by singular interactions.
Nevertheless we got an insight by solving the standard BCS
equation in the linearized form 
\begin{equation}
\label{bcst}
\Delta({\mbox{\boldmath $k$}})=
-\frac{1}{N_s}\sum_{{\mbox{\boldmath $p$}}}\Gamma (
{\mbox{\boldmath $k$}}-{\mbox{\boldmath $p$}})
\frac{\tanh \frac{\epsilon_{{\mbox{\boldmath $p$}}}}{2T}}
{2\epsilon_{{\mbox{\boldmath $p$}}}}\Delta({\mbox{\boldmath $p$}})
\end{equation}
and obtained $T_c$ vs $\kappa^2$ in the 
proximity of the ICDW instability.
$\epsilon^2({\mbox{\boldmath $p$}})=
\xi_{{\mbox{\boldmath $p$}}}^{2}+\Delta_{{\mbox{\boldmath $p$}}}^{2}$ 
with $\xi_{{\mbox{\boldmath $p$}}}$ being the electronic dispersion 
measured with respect to the Fermi energy $E_F$. $N_s$ is
the number of sites.
We considered a tight-binding model with hopping up to the fifth 
nearest neighbors 
 according to Ref. \onlinecite{BiSCO}. The used 
 parameters are appropriate for the band structure of the
BiSCCO compounds, 
giving an open Fermi surface and a van Hove singularity (VHS) 
slightly below the Fermi level (for electrons).
At optimal doping $\delta=0.17$
the value of $E_{F}=-0.1305$eV
is fixed to get the proper distance of the Fermi surface from the VHS 
($E_F - E_{VHS}=35$meV as determined experimentally).
The full bandwidth $W$ is $1.4$eV. 

We have verified that the $d$-wave transition has 
always a substantially larger critical temperature 
than the $s$-wave in the proximity the ICDW-QCP.
This is a consequence of the form of
Eq. (\ref{fitgamlr}), which, together with a
constant repulsion $\tilde{U}$, has a strongly
$q$-dependent attraction generically peaked at
rather small $q$'s.
Roughly, the $d$-wave  becomes favorable since the
average repulsion felt by the $s$-wave paired electrons 
exceeds the loss in condensation energy due to
the vanishing of the order parameter along the nodal regions.
Among the $d$ waves, the $d_{x^2-y^2}$ is preferred because 
the nodes occur in regions with small density of states.

As one can see in the inset of Fig. 4,
the BCS superconducting critical temperature
shows a strong increase  upon decreasing $\kappa^2$.
The actual behavior of $T_c$ is then obtained by
introducing the doping and temperature dependence
of $\kappa^2\equiv \kappa^2(\delta-\delta_c,T)$.
An additional (less important) doping dependence is due to 
the variation of the chemical potential with respect
to the VHS. In the quantum disordered phase
$T_c [\simeq T_c(\kappa^2(\delta-\delta_c,T=0))$]  will 
rapidly increase by decreasing doping towards $\delta_c$.
At a given doping $\tilde{\delta}\gtrsim \delta_c$ the 
BCS superconducting temperature will reach the
crossover line $\tilde{T}$ separating the quantum disordered from
the classical gaussian region.
In this latter region $\kappa^2$ weakly depends on doping
and a plateau in $T_c$ is reached.

To make the analysis of the over- and optimally doped
regions more quantitative we proceeded as follows.
First the value of the zero temperature coefficient $a$
in Eq.(\ref{kmax}) is extracted from the $\delta$-dependence of
$\Gamma(q,\omega=0)$ given in Ref.\onlinecite{CDG}. 
For the chosen values of the microscopic parameters
an optimal critical temperature 
$\tilde{T_c}\approx 90$K was put
in correspondence with a $\tilde{\kappa}^2 \approx 0.1$
(see inset of Fig. 4)
giving a coherence length of about 3 lattice 
units at the overdoped-optimal doping crossover. 
It was then possible to estimate the coefficient $b$
in Eq. (\ref{kmax}) from the relation 
$\tilde{\kappa}^2\simeq a(\tilde{\delta}-\delta_c)\simeq b\tilde{T_c}$,
which holds at the $\tilde{T}$ crossover line.
Once $a$ and $b$ are given, the 
$T_c$ vs $\kappa^2$ curves 
at the various fillings 
and the relation (\ref{kmax}) allow a complete determination
of $T_c=T_c[\delta,\kappa^2(\delta,T_c)]$  as a function of doping
in both the quantum disordered and the quantum critical regions.
The result is plotted in Fig. 4.
\begin{figure}
\vspace{-1 truecm}
\centerline{\hbox{\psfig{figure=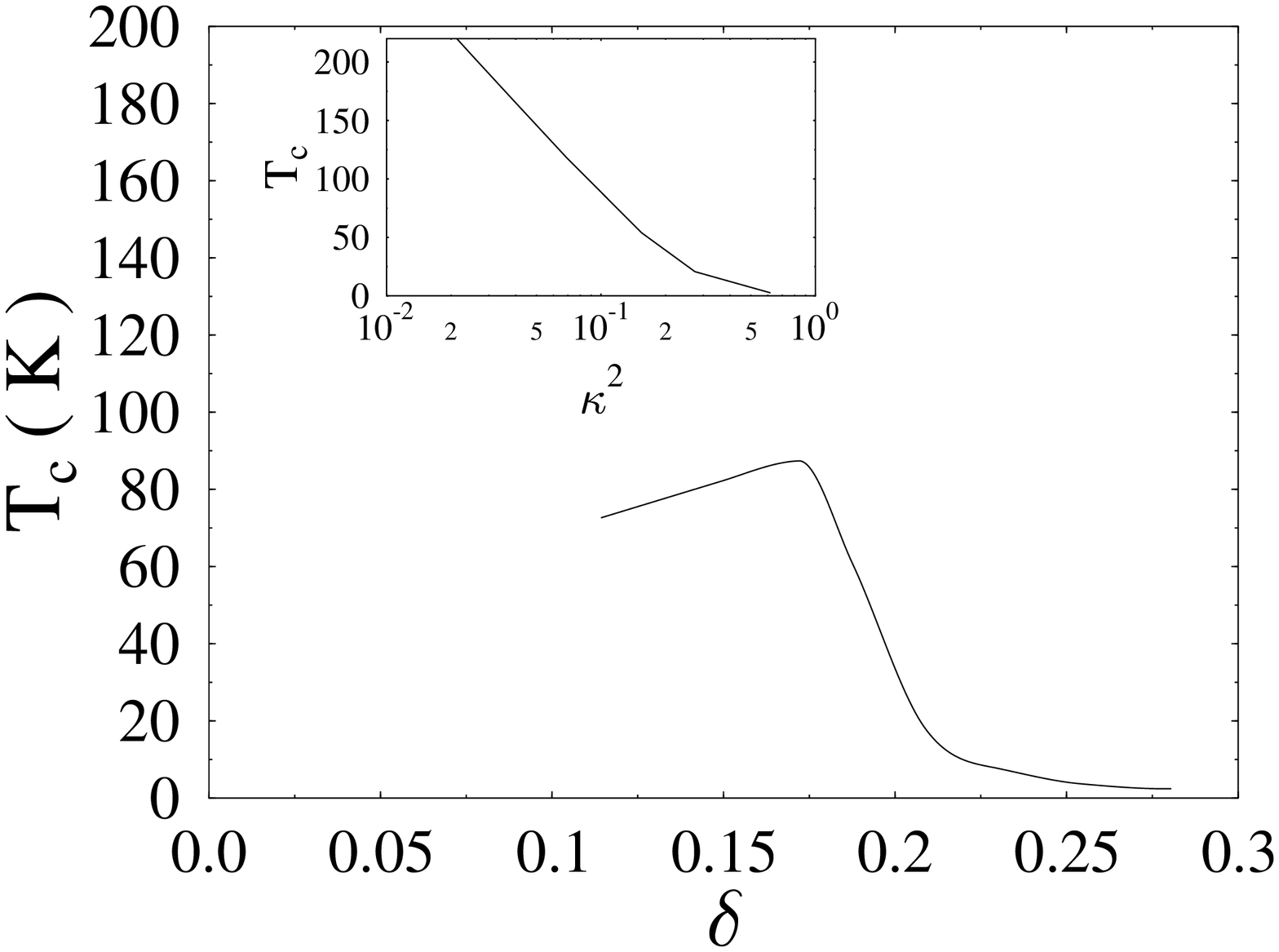,width=6.cm,angle=-90}}}
\vspace{-1 truecm}
{\small FIG. 4: BCS superconducting critical temperature
in the overdoped and optimally doped regions. The critical 
doping is $\delta_c\approx 0.17$, and 
$\tilde{\kappa}^2=0.1$ and $a=10$ are used
to estimate the crossover position (see text). 
In the pairing potential $\tilde{U}=0.2$ eV and $V=0.45$ eV
are used. Inset:
Dependence of the critical temperature on the mass parameter
$\kappa^2$ with $\delta=0.17$. }
\label{FIG4}
\end{figure}
The maximal critical temperature is obtained in correspondence
to the quantum-disordered/quantum-critical crossover
($\delta_{opt}\approx \tilde{\delta}$) \cite{notatopt}.
The slow decrease of $T_c$ by decreasing doping
in the quantum critical region is a consequence
of the decrease in the density of states.
Of course this is only an estimate depending on
the use of a weak coupling BCS scheme and of a model dependent evaluation
of $\kappa^2 (\delta-\delta_c, T)$.
Nevertheless, we remark that the 
experimentally observed rather rapid variation
of $T_c$ with doping in the overdoped region and the plateau
near optimal doping are quite naturally captured by our
description.

Notice that in discussing $T_c$ vs doping we have assumed that the main 
doping dependence is via $\kappa^2$. Indeed we have verified that the 
variations induced by moving $E_F$ are less relevant and calculating
$T_c$ at fixed $\kappa^2$,
the greatest values are obtained for $E_F\simeq E_{VHS}$. Then, a
finite $T_c$ would extend from very 
small (or even negative) doping up to
very high doping ($\delta > 0.6$). 
Strong variations of $T_c$ with doping, like those observed
in many cuprates, are hardly obtained in terms of a dependence on 
band parameters only (specifically, tuning the VHS). 
They are instead quite natural 
in the context of proximity to an instability, 
where doping controls the 
effective potential itself and not only the density of states.
This agrees with 
the experimental finding that at the maximum $T_c$ 
the VHS is not at the 
Fermi energy but below it \cite{shenreview}.

All the above analysis is confined to the 
over- and optimally doped regions of the phase
diagram where the fluctuations are not
strong enough to destroy the homogeneous 
character of the system. On the other hand the region on
the left of the mean-field critical curve for the ICDW
transition is characterized by strong thermal
fluctuations leading to ICDW precursors.
The ICDW fluctuations in the underdoped region
become critical in the proximity of the line $T_{CDW}(\delta)$
where the ICDW transition would occur in the absence of
superconducting pairing. 
Approaching $T_{CDW}$ the attractive fluctuations would lead to the 
formation of (local) pairs at the
curve $T^*(\delta)$. As a consequence of strongly
paired quasiparticles, below $T^*(\delta)$
pseudogap effects will show up 
as seen in many experiments (NMR, ARPES, 
optical conductivity, specific
heat, ...). However, despite
the strong pairing, the true superconducting critical
temperature is lower than $T^*$ and it
decreases inside the underdoped psudogap region.
This occurrence is schematically depicted in Fig. 3 by the
bifurcation of the heavy line. 
The idea of locally paired fermions
without long-range coherent superconductivity is a long-standing
recurrent concept in the context of high temperature superconductors,
where the coherence length is quite small and lead to
numerous investigations of mixed fermion-boson models
\cite{ranninger},
and Bose condensation vs BCS crossovers \cite{sademelo}.
The superconducting pseudogap problem has also been recently
analyzed for a single metallic stripe in an AF environment
\cite{EKZ}.

A simple model providing local pair formation is
the negative-U Hubbard model in the large-U regime,
with a  critical temperature for coherent
superconductivity, which decreases with increasing $\vert U\vert$. 
To apply this model to the underdoped cuprates, one
should then assume that the pairing potential U strongly increases
by decreasing the doping. However, in this case also the zero
temperature charge excitation gap would increase by the same amount
upon decreasing the doping (like $T^*$).
This is contrary to the observation \cite{loram,harris,campuzano}
that the low-temperature gap in the underdoped cuprates
weakly depends on doping, while $T^*$ increases fast by
decreasing $\delta$. 
We believe that this peculiar behavior requires
a remarkable temperature dependence
of the pairing potential as implied by our ICDW scenario
in the underdoped region. Although a full theory
of this complex phenomenon is far from being available, 
to get an insight on the physics of the pseudogap phase,
we introduce in $\kappa^2$ in
Eq. (\ref{fitgamlr}) a modified temperature dependence 
[with respect to Eq.(\ref{kmax})]
given by the distance from the critical line $T_{CDW}$ and add
a contribution due to the presence of a local superconducting gap
$\kappa^2 \equiv Max \left[ \vert \Delta_{Max}(T)\vert, 
c(T-T_{CDW}) \right]$. With this we aim 
to introduce in the pairing potential 
the stabilizing effect of the local superconducting order
on the ICDW instability. $\Delta_{Max}$ is
the maximum value of $\Delta({\mbox{\boldmath $k$}})$
in $k$-space.
The decreasing of the 
fluctuations above the ICDW critical line is considered
via $c(T-T_{CDW} )$. Of course $\kappa^2$
is selfconsistently related to the superconducting gap
determined by the potential itself. 
For simplicity, we assume again the BCS equation to be valid 
to estimate the (local) $d$-wave gap function. Then,
inserting the effective interaction (\ref{fitgamlr}) 
in the BCS equation, 
we obtained the behavior of $\Delta({\mbox{\boldmath $k$}})$
as a function of $T$ \cite{CDGP}. The result for $\Delta_{Max}$,
is reported in Fig. 5. Despite the oversimplified form 
of $\kappa^2$, that we have been using, the $T$ behavior
of $\Delta_{Max}$ bears a striking resemblance with
the analogous quantity recently measured with ARPES in underdoped
Bi-2212 samples \cite{harris}.
\begin{figure}
\vspace{-1 truecm}
\centerline{\hbox{\psfig{figure=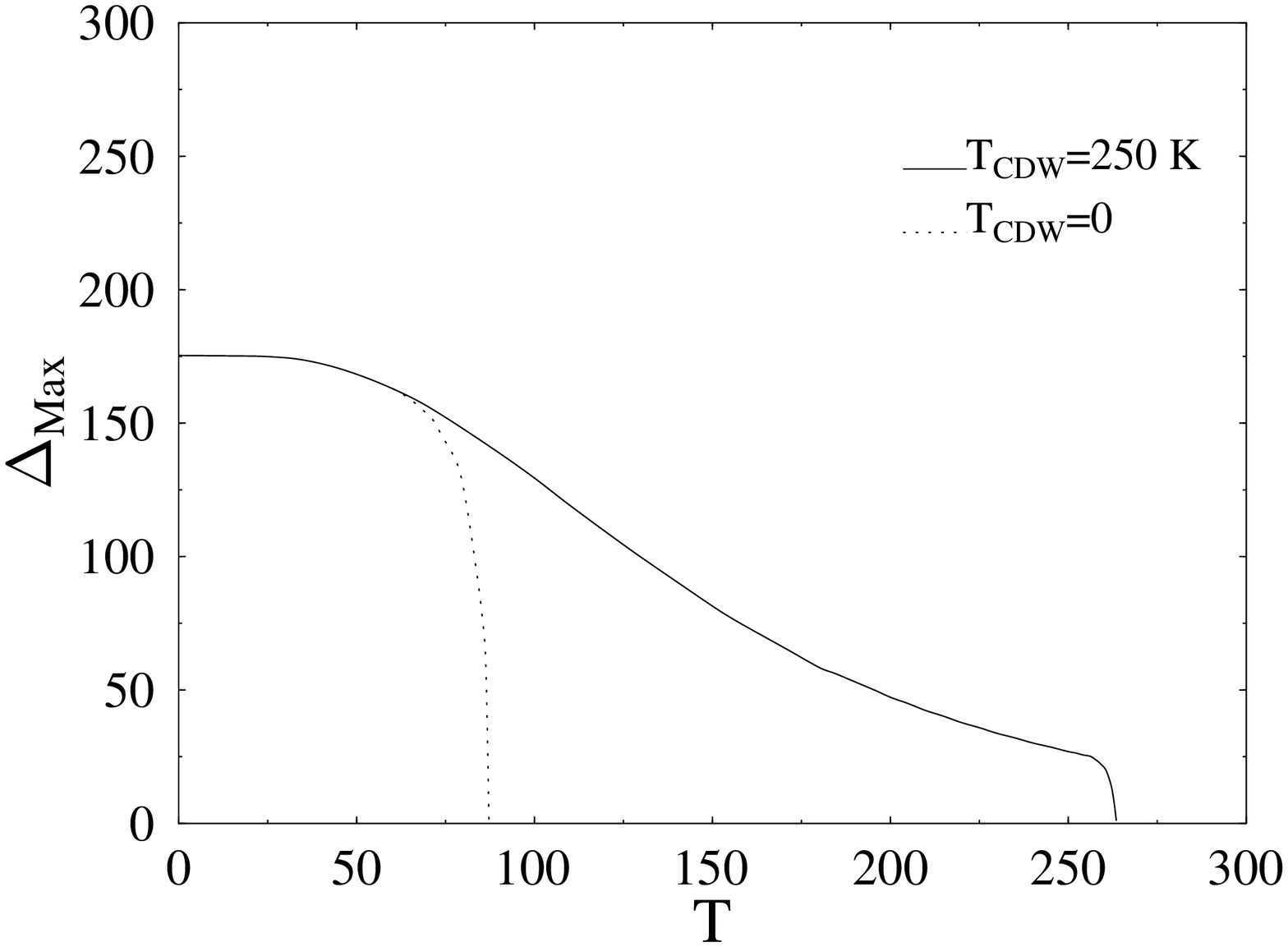,width=6.cm,angle=-90}}}
\vspace{-1 truecm}
{\small FIG. 5: Temperature 
dependence of the maximum value of the $d$-wave BCS superconducting 
gap in the underdoped region for different values of $T_{CDW}$.
In the pairing potential $V=0.45$ eV is used.}
\label{FIG5}
\end{figure}

Of course the above BCS treatment only deals with the
amplitude of the gap and says nothing on the
way a true superconducting phase coherence 
is established below a critical temperature $T_c<T^*$. 
For this we have to invoke phase fluctuations as in the
usual (large) negative-U Hubbard model. Here the strongly
peaked $q$-dependence of the pairing potential,
which only couples few points on the Fermi surface,
and precursors of (1d) stripe formation are expected to
give rise to strong phase fluctuations.

It is also important to emphasize that the occurrence of
local pairing prevents the actual establishing of the
ICDW long-range order \cite{notaCCCDGR}, so that $T_{CDW}$ looses
its meaning of a true transition line, merely
indicating the area where pairing and strong
local-dynamical ICDW order selfconsistently interplay.

As far as magnetic properties are concerned, we notice that
the presence of fluctuating hole-poor and hole-rich stripes
favors the presence of (locally commensurate) antiferromagnetic
fluctuations in the hole-poor regions. This generically explains
why, despite the rapid spoiling of the long-range AF by doping,
magnetic features survive up to much larger (optimal) doping.
Irrespective of the dominant mechanism leading to ICDW formation,
magnetism also contributes to further expelling charge from the
hole-poor stripes easily leading to non-linear phenomena and
higher harmonics generation. The presence of locally commensurate
AF order embedded in incommensurate stripes would also reconcile 
NMR and neutron scattering experiments \cite{bpt}.

\section{Conclusions}
The scenario presented in this work was 
based on the existence of ICDW instabilities in strongly correlated
systems and on generic properties of QCP's
in quasi-twodimensional systems. These properties were
exploited to establish some consequences of the ICDW-QCP scenario,
like, e.g., the marked non-FL character of the optimally
doped quantum disordered region,
the $d$-wave symmetry of the superconducting
order and the peculiar doping dependence of the superconducting
critical temperature in the over- and optimally doped regions.
In these homogeneous regions rather simple
theoretical approaches put our conclusions on a rather firm
ground. On the other hand, the  theoretical 
treatment of the underdoped phase
is made difficult by the interplay of three different local fields
associated to the tendency towards three ordered phases
(magnetic, charge-ordered and superconducting).
This renders our scenario
less established thus calling for further confirmations.
It is however remarkable that the 
temperature dependence of the pairing potential
implied by the ICDW scenario accounts well for the
peculiar (so far unexplained) temperature behavior of the local 
gap and for the non-trivial relation between $T^*$ and
$\Delta_{Max}(T=0)$.

%
\acknowledgments 
Part of this work was carried out with the financial support
of the Istituto Nazionale di Fisica della Materia, Progetto
di Ricerca Avanzata 1996.

%
%

\end{multicols}

\begin{references}
\bibitem{notarhoc}
The non-metallic behavior of $\rho_c$ could be due to the fact
that the ICDW fluctuations in the planes are enhanced below 
$T_0$ rendering difficult the coherent single-particle hopping.
\bibitem{notatopt} 
The fact that the highest $T_c$ is obtained in correspondence
to the crossover line
between the quantum-critical and the quantum disordered
regions is a consequence of the simplified form of $\kappa^2$
in Eq.(\ref{kmax}).
\bibitem{notaCCCDGR} 
Indeed it was shown in a Kondo-lattice-like
model that superconductivity stabilizes phase separation already
at mean field level\cite{CCCDGR}.
\end{references}
\end{document}